\documentclass[conference]{IEEEtran}
\IEEEoverridecommandlockouts
\usepackage{amsmath,amssymb,amsfonts}
\usepackage{algorithmic}
\usepackage{graphicx}
\usepackage{parskip}
\usepackage{textcomp}
\usepackage{mdframed}
\usepackage{xcolor}
\usepackage[
backend=biber,
style=numeric,
citestyle=numeric
]{biblatex}
\usepackage[tight,footnotesize]{subfigure}
\begin{document}

\title{An empirical investigation into audio pipeline approaches for classifying bird species\\
}

\author{
\IEEEauthorblockN{David Behr}
\IEEEauthorblockA{\textit{Department of Computer Science} \\
\textit{University of Pretoria}\\
Pretoria, South Africa \\
u14032512@tuks.co.za}
\and
\IEEEauthorblockN{Ciira wa Maina}
\IEEEauthorblockA{\textit{Department of Electrical and Electronic Engineering} \\
\textit{Dedan Kimathi University of Technology}\\
Nyeri, Kenya \\
ciira.maina@dkut.ac.ke}
\and
\IEEEauthorblockN{Vukosi Marivate}
\IEEEauthorblockA{\textit{Department of Computer Science} \\
\textit{University of Pretoria}\\
Pretoria, South Africa \\
vukosi.marivate@cs.up.ac.za}}

\maketitle

\begin{abstract}
This paper is an investigation into aspects of an audio classification pipeline that will be appropriate for the monitoring of bird species on edges devices. These aspects include transfer learning, data augmentation and model optimization.The hope is that the resulting models will be good candidates to deploy on edge devices to monitor bird populations. Two classification approaches will be taken into consideration, one which explores the effectiveness of a traditional Deep Neural Network(DNN) and another that makes use of Convolutional layers.This study aims to contribute empirical evidence of the merits and demerits of each approach.

\end{abstract}


\section{Introduction}
Biodiversity faces a number of threats due to a host of reasons, ranging  from human encroachment to climate change. This study focuses on the biodiversity monitoring of bird species. Unfortunately current approaches of monitoring bird species such as conducting manual surveys are flawed in many ways. These techniques are prune to mis-identification, are expensive and are temporal in nature.

Machine Learning techniques provide better solutions that enable real time monitoring and reduce cost significantly.The combination of traditional digital signal processing and Neural Networks such as Convolutional Neural Networks(CNN's) have proven to be effective in environmental sound recognition (ESR), and these techniques can be applied to biodiversity monitoring.

The use of machine learning techniques present a few obstacles such as the requirement for a large amount of data to train on, something that is lacking with some bird species. However there are techniques such as data augmentation and transfer learning that can be used to overcome this.

The focus of this paper will be the exploration of two possible classification pipelines that rely on different machine learning architectures to learn. Namely a Dense Neural Network (DNN) and a Convolutional Neural Network (CNN) will be investigated in depth. Both models will make use of data augmentation to overcome dataset limitations, however transfer learning will only be applied to the CNN. This is due to the fact that CNN's contain transferable features.

Finally both model types will undergo model optimization and compression in order to explore the possibility of deploying these machine learning models on edge devices.

\section{Methodology}
This section discuses details such as parameter choices pertinent to each experiment and model.

\subsection{Pre-processing}
The point of pre-processing is to generate a feature set that is adequate for the networks to train on.

Parameters used to generate the Mel-spectrograms:
\begin{enumerate}
    \item sampling rate = 44100
    \item hop\_length = 348* duration
    \item fmin = 20
    \item fmax = sampling\_rate //5
    \item n\_mels = 128
    \item n\_fft = n\_mels * 20
    \item samples = sampling\_rate * duration
\end{enumerate}

When applying the above parameters to a sound of any length a Mel-Spectrogram of the same dimension is generated. Thus, using the whole sound is possible.

\subsection{data augmentation}
In the experiments conducted for this paper, data augmentation on the Mel Spectrogram is reserved for the CNN pipeline whereas augmentation performed on the audio is reserved for the DNN pipeline. Each data augmentation technique is applied at random at a set rate only on training data. Validation data and test data are not augmented in any way. In order to limit the scope of the this research,combinations of data augmentation techniques will not be applied. Augmentation techniques that will be applied to the Mel-spectrograms include \emph{mixup}, \emph{cutmix} and \emph{specAug}.  Augmentation techniques that will be applied to the audio includes pitch shift, noise injection, time shift and speed shift.

\subsubsection{Datasets}
The datasets that will used in this study are derived from the \textit{Xeno-Canto} ~\cite{xeno} repository. The main dataset contains 24000 samples of bird species from mainly north America. 
The second dataset includes bird species from southern africa and the third dataset includes endagered bird species from south africa constructred using the \textit{Xeno-Canto} api and IUCN (International Union for Conservation of Nature)  redlist 

\subsection{Model Architecture and Hyper Paramater Tuning}
The hyper parameters of the baseline convolutional models and deep neural networks used in this paper were tuned via an automatic hyper parameter tuning technique called Hyperband Search \cite{li2017hyperband}.In the case of the baseline convolutional model, a semi structured model based on the VGG (Visual Geometry Group) architecture was passed into the hyperband algorithm to determine which loss function, learning rate and layer configuration worked best for the given dataset.

\subsection{Transfer Learning}
This paper will assess the success of 4 pre-trained networks against a baseline, all of the pre-trained networks listed have been trained on the popular Imagenet dataset. The networks are: VGG16 \cite{simonyan2014very}, VGG19 \cite{simonyan2014very}, MobileNetV2 \cite{howard2017mobilenets}, InceptionResnetV2 \cite{szegedy2016inception}. These base CNN Networks are available through the popular Keras Library

The Transfer Learning flow that will be used in this paper is as follows:
\begin{itemize}
    \item Freezing the convolutional base and attaching a dataset specific classifier.
    \item Training the attached classifier for a few epochs.
    \item Unfreezing the convolutional base and fine-tuning the whole network for a few epochs.
\end{itemize}

\subsection{Model Compression and Optimization}
Model Compression and Optimization are critical for models that will potentially be deployed on edge devices, which is the case with the models developed in this study. Compression and Optimization techniques include pruning, quantization and weight clustering. For the Purposes of this study pruning and quantization will be explored, independently and together. In addition to post training pruning and quantization, quantization aware training discussed in \cite{han2015deep} as online training will also be trialed.

\section{Results}
This section provides an in depth analysis of the results of each approach taken to classify birds and how these approaches perform on new data.

\subsection{CNN Pipeline}
\begin{figure}[h]
    \centering
    \includegraphics[width=0.6\columnwidth]{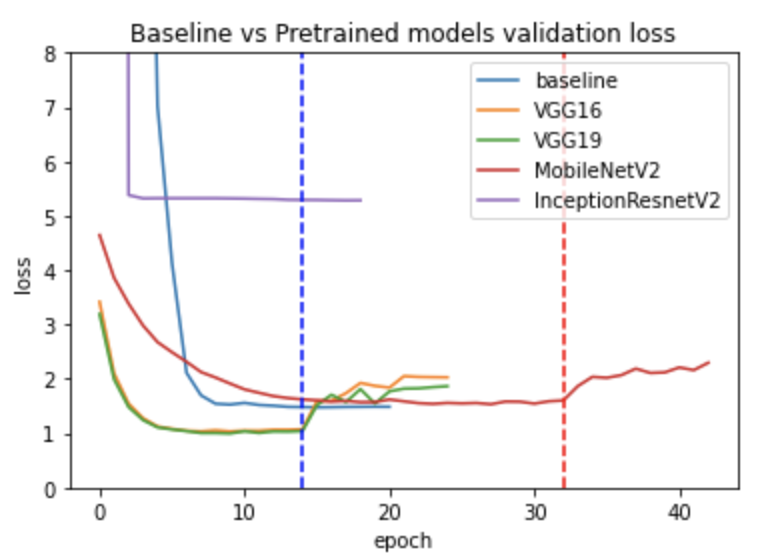}
    \caption{baseline vs pre-trained networks}
    \label{fig:baseline-vs-pretrained}
\end{figure}

The best performing models are shown in Table \ref{tab:cnn-baseline-models}, each of which were trained using the different data augmentation techniques that were mentioned in the methodology. The performance of the models were confirmed via 5 k cross fold validation. All of the pre-trained models besides InceptionResnetv2 outperform the baseline. Based on the precision metric, VGG16 and VGG19 deliver the best results and MobileNetv2 produces similar results with a smaller architecture. Figure \ref{fig:baseline-vs-pretrained} shows the validation accuracies for the CNN models. The dotted lines indicate the point at which fine tuning of the baseline began. It is interesting to note that fine tuning tended to have a negative affect on the performance of the model since it increased over fitting. 

\begin{table}[h]
\centering
\caption{Best Results for Each Model}
\label{tab:cnn-baseline-models}
\begin{tabular}{|l|l|l|l|l|l|}
\hline
\textit{}       & Precision & Recall & F1 Score & loss & accuracy \\ \hline
Baseline        & 0.81              & 0.75           & 0.75     & 1.37            & 0.78                \\ \hline
MobileNetv2     & 0.84              & 0.84           & 0.82     & 1.89            & 0.84                \\ \hline
VGG16           & 0.87              & 0.86           & 0.84     & 1.9             & 0.8                 \\ \hline
VGG19           & 0.86              & 0.85           & 0.83     & 0.96            & 0.85                \\ \hline
InceptionResnet & 0.012             & 0.01           & 0.009    & 5.1             & 0.0084              \\ \hline
\end{tabular}
\end{table}

InceptionResnetV2's inability to learn is quite strange, it could be attributed to its complex architecture. InceptionResnet was developed to perform well on image classification tasks that may include objects of interest at various sizes and locations, therefore each inception block contains a few kernel sizes that each convolve an image and the results are then merged. Since the information in a Mel-spectrogram is globally distributed it may be difficult for the architecture to consider the whole image as a location of interest.

In terms of the baseline model, over 150 classes out of the 205 classes have perfect recall and precision. Some of the species that have perfect recall and precision include the Alder-flycatcher and the Great-Crested-Flycatcher. Species that have poor recall include the Red Eyed Vireo and the the pileated Woodpecker. It is not clear from the spectrograms as to why some species are easier to classify.

\begin{figure}[h]
    \centering
    \subfigure{\includegraphics[width=0.24\textwidth]{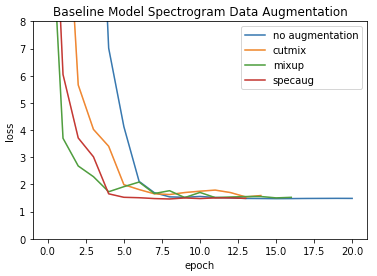}} 
    \subfigure{\includegraphics[width=0.24\textwidth]{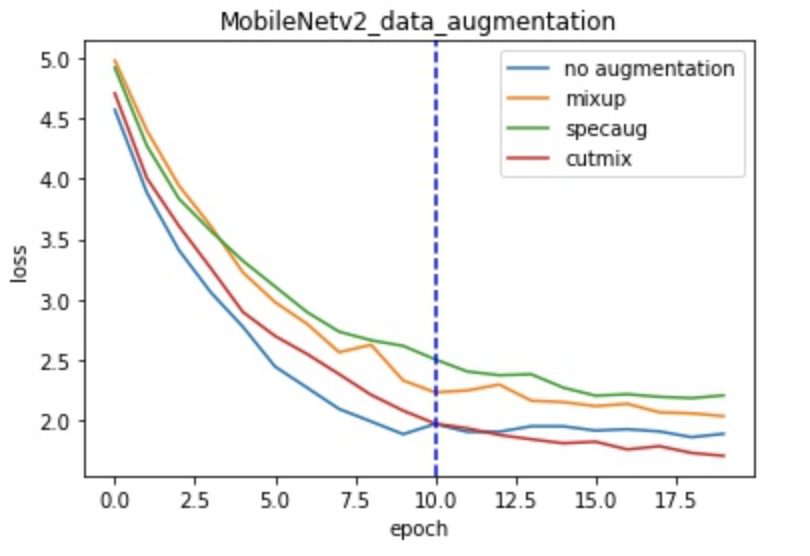}}
    \caption{(a) Baseline DNN confusion matrix (b) Audio Feature Data Augmentation}
    \label{fig:cnn-data-aug}
\end{figure}

Data Augmentation techniques such as mixup \cite{Mixup}, cutmix \cite{Cutmix} and specaug \cite{Specaug} had little to no effect on the the pretrained models as seen in table \ref{tab:data-aug-data}. In the case of MobileNetV2, data augmentation tended to have a negative effect on all metrics such as precision and recall as seen in figure \ref{fig:cnn-data-aug} b. However evaluation loss did improve when applying cutmix to the MobileNetV2 model. \textit{Specaug} slightly increased precision in the VGG16 model, however not by enough to justify its application. Cutmix slightly improved precision and loss when applied to the VGG19 model, however this slight change does not justify its use.

The application of data augmentation to the InceptionResnetV2 was not able to increase the models performance. Similar results were observed when applying data augmentation to the baseline model, it was also noted that occasionally data augmentation techniques actually increased model loss as seen in figure \ref{fig:cnn-data-aug}. Overall Specaug was able to increase precision in two models, however there is no consensus on which data augmentation technique is best as the results differ depending on which model is used.

An interesting observation is that each data augmentation application changes the dominance of recall and precision of some species. For example when applying data augmentation to MobileNetworkV2 the Finch and Barbets precision increased where as the Flycatchers recall and precision decreased. This indicates that applying class conditional augmentation may benefit performance. A similar observation was made by the authors of \cite{salamon2017deep} when applying data augmentation techniques to environmental sounds.

\begin{table}[]
\centering
\caption{Data Augmentation results from CNN models}
\label{tab:data-aug-data}
\begin{tabular}{|l|l|l|l|l|l|}
\hline
\textit{}       & precision & recall & F1 score & loss   & accuracy \\ \hline
\multicolumn{6}{|c|}{\textbf{Baseline Data Augmentation}}           \\ \hline
No augmentation & 0.79      & 0.76   & 0.75     & 1.3784 & 0.77     \\ \hline
cutmix          & 0.76      & 0.73   & 0.71     & 1.57   & 0.73     \\ \hline
Specaug         & 0.81      & 0.75   & 0.75     & 1.37   & 0.75     \\ \hline
Mixup           & 0.79      & 0.79   & 0.75     & 1.44   & 0.77     \\ \hline
\multicolumn{6}{|c|}{\textbf{VGG16 Data augmentation}}              \\ \hline
No augmentation & 0.86      & 0.85   & 0.84     & 1.79   & 0.85     \\ \hline
cutmix          & 0.86      & 0.85   & 0.84     & 1.00   & 0.85     \\ \hline
Specaug         & 0.87      & 0.86   & 0.84     & 1.9    & 0.8      \\ \hline
mixup           & 0.86      & 0.85   & 0.84     & 1.1    & 0.83     \\ \hline
\multicolumn{6}{|c|}{\textbf{VGG19 Data augmentation}}              \\ \hline
No augmentation & 0.84      & 0.85   & 0.83     & 1.9    & 0.85     \\ \hline
cutmix          & 0.86      & 0.85   & 0.83     & 0.96   & 0.85     \\ \hline
Specaug         & 0.85      & 0.85   & 0.83     & 1.7    & 0.85     \\ \hline
Mixup           & 0.85      & 0.85   & 0.83     & 1.00   & 0.85     \\ \hline
\multicolumn{6}{|c|}{\textbf{MobileNetV2 Data Augmentation}}        \\ \hline
No augmentation & 0.84      & 0.84   & 0.82     & 1.89   & 0.84     \\ \hline
cutmix          & 0.84      & 0.83   & 0.81     & 1.2    & 0.83     \\ \hline
Specaug         & 0.79      & 0.80   & 0.78     & 1.4    & 0.81     \\ \hline
Mixup           & 0.83      & 0.84   & 0.8      & 1.3    & 0.83     \\ \hline
\multicolumn{6}{|c|}{\textbf{InceptionResnetV2 Data Augmentation}}  \\ \hline
No augmentation & 0.0097    & 0.0097 & 0.0081   & 5.2    & 0.0081   \\ \hline
cutmix          & 0.0093    & 0.0099 & 0.0082   & 5.4    & 0.0079   \\ \hline
Specaug         & 0.0083    & 0.0083 & 0.0079   & 5.9    & 0.0073   \\ \hline
Mixup           & 0.012     & 0.01   & 0.009    & 5.1    & 0.0084   \\ \hline
\end{tabular}
\end{table}

The application of model optimization and compression techniques had a positive effect on the baseline as can be seen by the results recorded in table \ref{tab:baseline-model-optimization}. Baseline accuracy increased by 5\% after pruning the model to 80\% sparsity. This is surprising, since pruning generally has a negative affect on model accuracy. Pruning also increased recall and precision by approximately 4\%. Pruning reduced the baseline model size by approximately 96\% and the application of 8 bit quantization to the pruned model reduced the pruned model by an additional 75\%. This huge reduction in model size is ideal for deploying it on an edge device.

It also suggests that neural networks are typically over parameterized and there is significant redundancy in the neural network. \cite{denil2013predicting} shows that this inherent network redundancy is a waste of both computational resources and memory usage.
This unique situation in which accuracy increased was also noticed in \cite{blalock2020state} in which the authors compared a plethora of popular CNN architectures such as Resnet before and after pruning was performed.

\begin{figure}[h!]
    \centering
    \includegraphics[width=0.60\columnwidth]{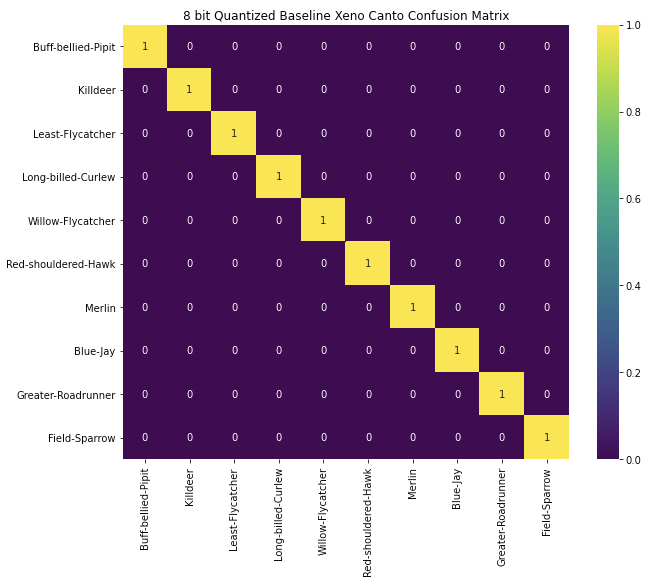}
    \caption{confusion matrix of model that has been pruned and quantized}
    \label{fig:pruned_quantized_baseline}
\end{figure}

The authors of \cite{han2015deep} show that the accuracy of a pruned network drops when the network is compressed 8\% below its original size; the same occurs with quantized networks. However when the two methods are combined the the network can be compressed to 3\% of its original size without losing the loss of accuracy. Similar results were noticed in the experiments conducted for this report; by combining 8 bit Quantization with pruning the baseline models accuracy increased by 13\%, precision and recall also increased by about 5\%. The confusion matrix in figure \ref{fig:pruned_quantized_baseline} shows that the 8 bit quantized model shows confidence in each of its classes and does not confuse any classes. The application of quantization aware training was not able to increase accuracy or any other performance metric obtained in previous experiments.

\begin{table}[h!]
\centering
\caption{Baseline Model Optimization \\ Where P stands for pruning, Q stands for Quantization and a.t stands for aware training}
\label{tab:baseline-model-optimization}
\begin{tabular}{|l|l|l|l|l|l|l|}
\hline
\textit{}                         & Precision & Recall & F1 Score & loss & acc & size \\ \hline
Baseline                          & 0.81      & 0.75   & 0.75     & 1.37 & 0.78     & 1.2G \\ \hline
P                           & 0.85      & 0.83   & 0.82     & 0 .96 & 0.83     & 376M \\ \hline
P and Q          & 0.90      & 0.90   & 0.90     & 0.80 & 0.93     & 94M  \\ \hline
8 bit Q                & 0.93      & 0.93   & 0.93     & 0.78 & 0.96     & 94M  \\ \hline
8 bit Q a.t & 0.89      & 0.90   & 0.89     &0.81 &  0.91    &    105M \\ \hline
\end{tabular}
\end{table}

\subsection{DNN Pipeline}

The following results were obtained from pairing audio features such as zero crossing rate with a Deep Neural Network. Principle component analysis was used to evaluate how much each feature contributes to variance . The initial principle components (PC's) contribute significantly to the variance, and it drops off towards the end. However, it is clear that each principle component contributes to the variance indicating that each PC is useful.

\begin{table}[h]
\centering
\caption{Baseline Audio Feature Model}
\label{tab:baseline-model-optimization-2}
\begin{tabular}{|l|l|l|l|l|l|l|}
\hline
\textit{} & Precision & Recall & F1 Score & loss & acc & training time \\ \hline
Baseline  & 0.81      & 0.80   & 0.78     & 1.59 & 0.79     & 3.29 secs     \\ \hline
\end{tabular}
\end{table}

The baseline neural network was able to deliver impress classification metrics, however the baseline CNN's performance is still superior. Similar results were obtained by the authors of \cite{hershey2017cnn}. DNN's are not able to benefit from the application of transfer learning due to the nature of the feature dependent weights, these weights do not share simple properties such as Gabor filters in CNN's that can be transferred to other tasks. Regardless there are still some benefits to using a DNN that cannot be overlooked such as training time and small model size.

\begin{figure}[h]
    \centering
    \subfigure[a]{\includegraphics[width=0.4\textwidth]{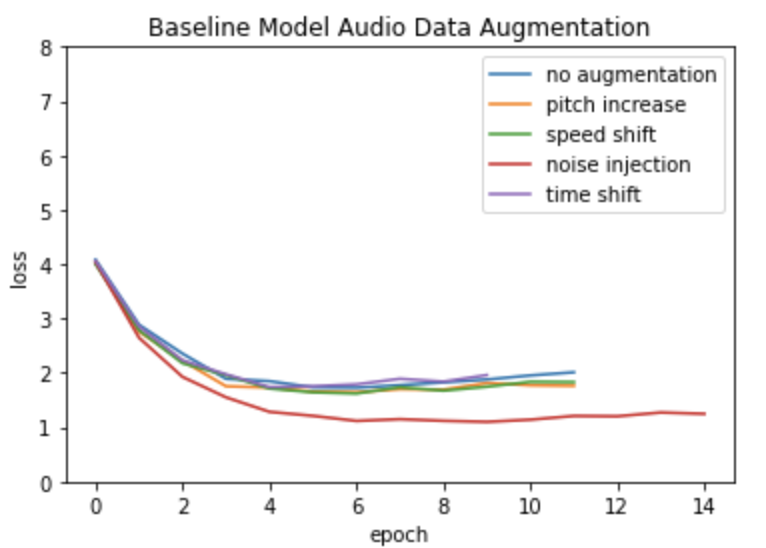}} 
    \caption{(a) Audio Feature Data Augmentation}
    \label{fig:foobar}
\end{figure}

\begin{table}[h]
\centering
\caption{Baseline Audio Feature Augmentation}
\label{tab:bl-audi-feat-aug}
\begin{tabular}{|l|l|l|l|l|l|}
\hline
\textit{}       & Precision & Recall & F1 Score & loss & accuracy \\ \hline
No augmentation & 0.81      & 0.80   & 0.78     & 1.59 & 0.79     \\ \hline
Noise Injection & 0.83      & 0.81   & 0.79     & 1.7  & 0.81     \\ \hline
Time Shift      & 0.72      & 0.72   & 0.70     & 1.7  & 0.72     \\ \hline
Pitch Shift     & 0.80      & 0.79   & 0.77     & 1.9  & 0.78     \\ \hline
Speed Shift     & 0.81      & 0.81   & 0.79     & 1.39 & 0.81     \\ \hline
\end{tabular}
\end{table}
The application of data augmentation techniques such as noise injection and speed shift to the baseline model are recorded in table \ref{tab:bl-audi-feat-aug}.  It is clear that most of the techniques have little to no effect. Noise injection and speed shift are the only data augmentation techniques that were able to significantly increase all performance metrics besides loss. The application of noise injection to the DNN was able to outperform the baseline CNN's performance.

The change in recall and precision metrics for different species was also noticed when applying different data augmentation techniques to the DNN. This reinforces the idea that conditional class based augmentation may be beneficial.

\begin{figure}[h]
    \centering
    \includegraphics[width=7cm]{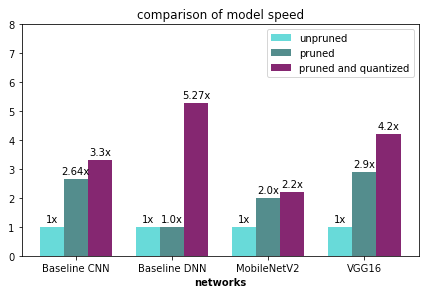}
    \caption{Compared with the original baseline CNN network, the pruned baseline network achieves a 2.64x speedup on the CPU and the pruned and the quantized network and pruned network achieves a 3.3x speed up on the Baseline}
    \label{fig:model_speedup}
\end{figure}

Pruning the DNN reduces the the model size by 65\% as seen in table \ref{tab:zindi-cnn-optimized} and increase the accuracy  precision and recall by approximately 3\% as seen in table \ref{tab:baseline-model-optimization-dnn} . The application of quantization to the pruned model has less of an effect on model size reduction in comparison to the size change observed with the baseline CNN. The application of quantization also decreases accuracy and precision. Applying 8 bit quantization alone without pruning produced similar results to pruning.

Figure \ref{fig:model_speedup} shows inference speed up for the best performing models. the results have been normalized to the baseline. It it clear that the application of pruning causes significant reduction of model inference times. The DNN model sees the greatest speedup of 5.2x due to the application of pruning and quantization. This is significant because it could potentially classify bird species in real time if the classification performance of the model increased. The results noted in \ref{fig:model_speedup}  make sense, since quantization decreases computational overhead since computations on 8 bit numbers are much faster in comparison to FP32 arithmetic.

\begin{table}[h]
\centering
\caption{DNN Optimized Model }
\label{tab:zindi-cnn-optimized}
\begin{tabular}{|l|l|l|l|l|l|l|}
\hline
\textit{}                & Precision & Recall & F1 Score & loss & accuracy & size \\ \hline
Baseline                 & 0.32      & 0.32   & 0.29     & 3.23  & 0.32     & 8.5M \\ \hline
\end{tabular}
\end{table}

\begin{table}[h]
\centering
\caption{Baseline DNN Model Optimization \\ Where P stands for Pruning and Q stands for Quantization}
\label{tab:baseline-model-optimization-dnn}
\begin{tabular}{|l|l|l|l|l|l|l|}
\hline
\textit{}                         & Precision & Recall & F1 Score & loss & acc & size\\ \hline
Baseline                          & 0.32      & 0.32   & 0.29     & 3.23 & 0.32     & 8.5M \\ \hline
P                           & 0.35      & 0.34   & 0.30     & 3.00 & 0.35     & 2.9M \\ \hline
P and Q          & 0.34      & 0.32   & 0.29     & 3.12 & 0.33     & 2.7M  \\ \hline
8 bit Q                & 0.35      & 0.29   & 0.30     & 2.99 & 0.34     & 2.9M  \\ \hline
\end{tabular}
\end{table}

\subsection{Comparison of Pipelines trained on new data}

This section explores the results that were obtained from applying the the best models from the previous two sections to new data.

\begin{figure}[h]
    \centering
    \includegraphics[width=7cm]{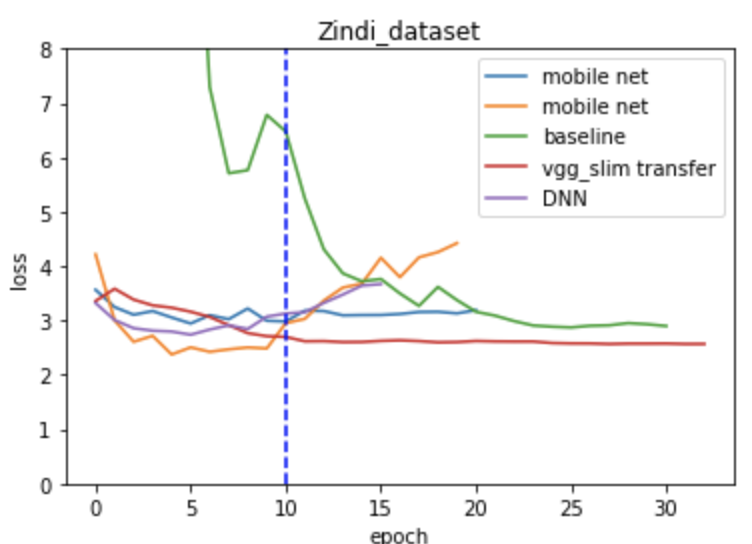}
    \caption{Models Trained on South African Bird Data}
    \label{fig:zindi-model-comparison}
\end{figure}

\begin{figure}[h]
    \centering
    \includegraphics[width=0.60\columnwidth]{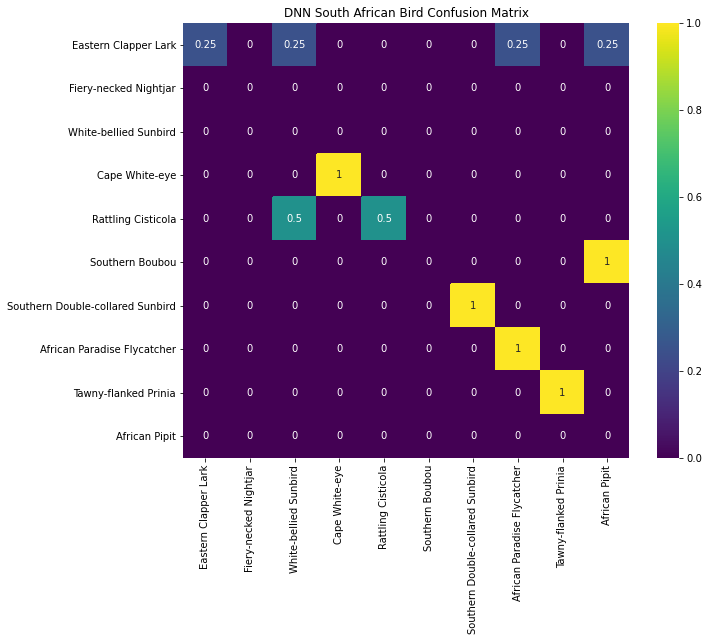}
    \caption{DNN trained on South African bird data}
    \label{fig:baseline_zindi_CM}
\end{figure}

\begin{figure}[h]
    \centering
    \includegraphics[width=0.60\columnwidth]{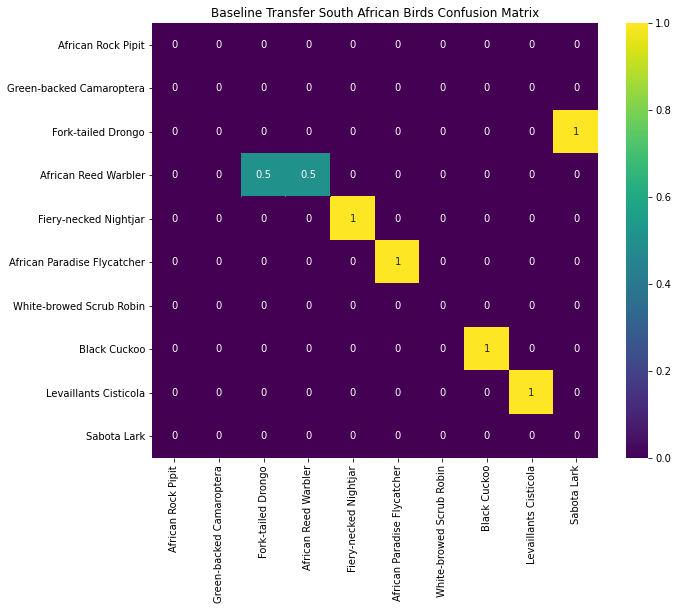}
    \caption{Transfer performance DNNs trained on South African bird data}
    \label{fig:baseline_transfer_cm}
\end{figure}

\begin{table}[h]
\centering
\caption{CNN pipeline trained with South African Bird data}
\label{tab:spec-zindi-compare}
\begin{tabular}{|l|l|l|l|l|l|}
\hline
\textit{}         & Precision & Recall & F1 Score & loss   & accuracy \\ \hline
Baseline          & 0.32      & 0.32   & 0.29     & 3.23   & 0.32     \\ \hline
VGG16             & 0.27      & 0.26   & 0.24     & 6.3    & 0.26     \\ \hline
VGG16\_transfer   & 0.29      & 0.279  & 0.29     & 3.6794 & 0.35     \\ \hline
Baseline transfer & 0.33      & 0.35   & 0.31     & 2.62   & 0.35     \\ \hline
\end{tabular}
\end{table}

The table \ref{tab:spec-zindi-compare} shows the results for the best performing CNN models from the previous sections trained on the new data. Training the baseline model proposed in the CNN section from scratch on the new data and using the pre-trained VGG16 model trained on the ImageNet dataset did not perform well. However the VGG16 model fine tuned on the \textit{Xeno-Canto} dataset performed better than the CNN baseline in terms of accuracy.

Using the baseline CNN as the baseline for transfer learning achieved the best results out of all the models. It had significantly lower loss and better precision and recall metrics. The confusion matrix in figure \ref{fig:baseline_transfer_cm} is limited to 10 of the 40 classes. It is clear that the model has high confidence in some of the classes such as the Black Cuckoo, but confuses confuses some classes for another. For example the model classifies the Fork tailed Drongo as a Sabota Lark with high confidence. There is also some confusion between the African reed warbler and the fork tailed drongo.

\begin{table}[h]
\centering
\caption{Audio Feature Pipeline trained on South African Bird Data}
\label{tab:audio-feat-zindi-compare}
\begin{tabular}{|l|l|l|l|l|l|}
\hline
\textit{} & Precision & Recall & F1 Score & loss & accuracy \\ \hline
Baseline  & 0.15      & 0.19   & 0.15     & 3.3  & 0.18     \\ \hline
\end{tabular}
\end{table}

The CNN models outperform the DNN model when applied to a new limited dataset, that has fewer samples for each class and the mean audio length is much less. It is clear that the accuracy, precision and recall achieved by the DNN pipeline are not desirable and make the model unsuitable for real world use regardless if the inference time and model size are desirable.

In terms of the CNN models, using the baseline model trained on \textit{Xeno-Canto} data performed the best in every metric. This indicates that the features learned in the baseline are transferable. Using the VGG16 model that was trained on \textit{Xeno-Canto} data achieved similar accuracy to the baseline transfer but was not able to achieve better accuracy or recall. Thus the best model to use for new dataset's is the baseline model trained on \textit{Xeno-Canto} data.

The confusion matrix in figure \ref{fig:baseline_zindi_CM} shows that the DNN model has high confidence in some classes and tends to classify a few correctly. However it shows a lot more uncertainty that the CNN model in figure  \ref{fig:baseline_transfer_cm}.
The Eastern Clapper Lark is confused with quite a few birds species such as the African Pipit and African Paradise Flycatcher. As with the initial models trained on the initial dataset, the models trained on the smaller south African birds dataset experience a gain in accuracy after pruning and quantization are performed, these results are tabulated in \ref{tab:zindi-cnn-optimized-2}.

\begin{table}[h]
\centering
\caption{South African Bird Species Optimized Model \\ Where P stands for Pruning and Q stands for Quantization}
\label{tab:zindi-cnn-optimized-2}
\begin{tabular}{|l|l|l|l|l|l|l|}
\hline
\textit{}                & Precision & Recall & F1 Score & loss & acc & size \\ \hline
Baseline                 & 0.32      & 0.32   & 0.29     & 3.23  & 0.32     & 1.2G \\ \hline
Pruning                  & 0.36      & 0.35   & 0.34     &2.1   & 0.38     &  340M    \\ \hline
P and Q & 0.38      & 0.36   & 0.35     &1.9   & 0.4      &  94M    \\ \hline
8 bit Q       & 0.36      & 0.37   & 0.35     &1.9   & 0.39       &  94M   \\ \hline
\end{tabular}
\end{table}

\subsection{Endangered Bird Species}

\begin{figure}[h]
    \centering
    \includegraphics[width=0.60\columnwidth]{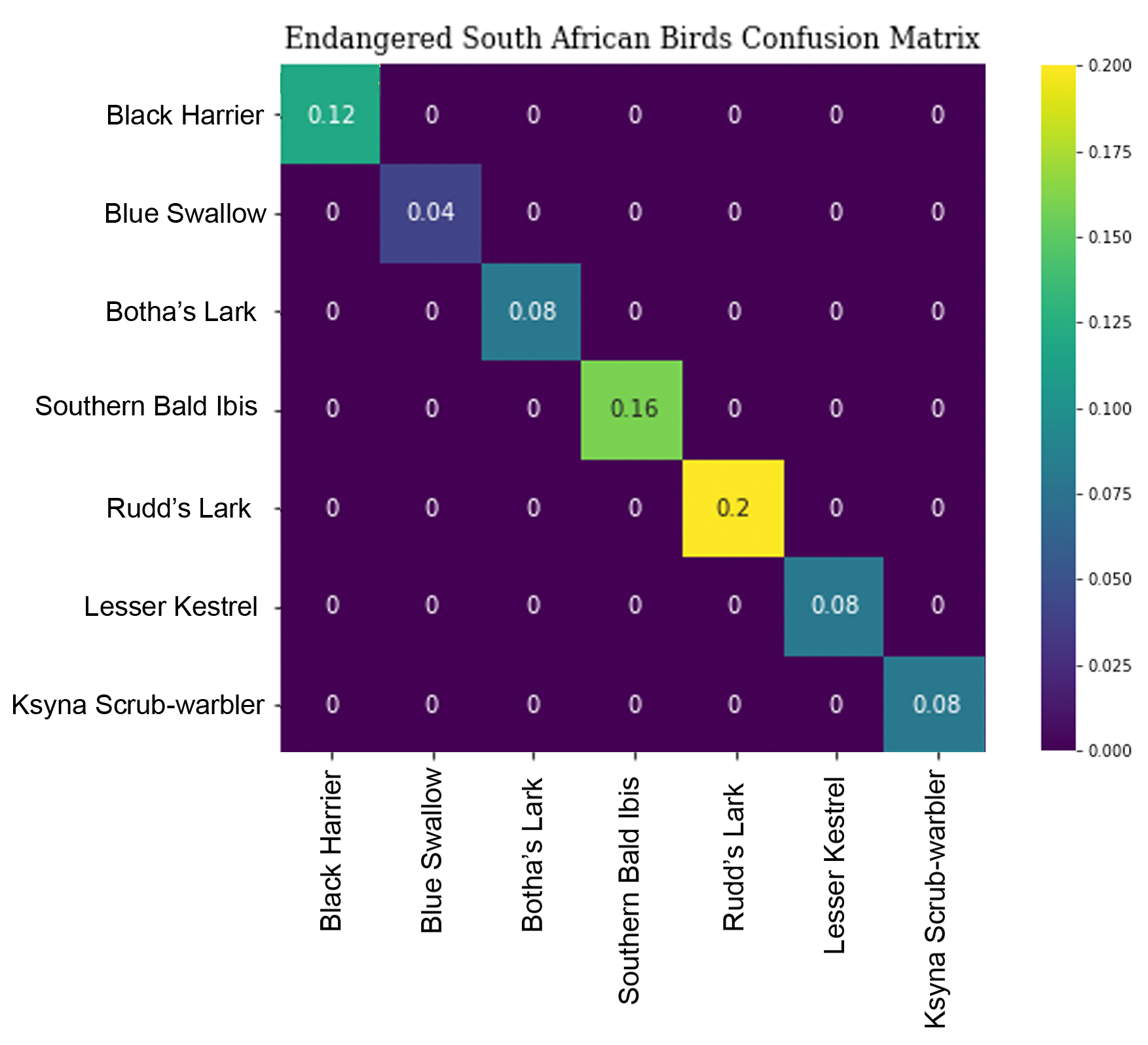}
    \caption{Baseline transfer applied to endangered species}
    \label{fig:endangeredCM}
\end{figure}
The monitoring of endangered bird species holds a lot of value, this section provides analysis of the application of the models mentioned in the previous sections when applied to endangered species.Using the baseline transfer learning model to classify a small set of endangered birds in southern Africa that are listed on the IUCN Red list yields promising results. Using a dataset of 7 endangered Southern African species compiled using the \textit{Xeno-Canto} API.

It is clear in the confusion matrix in figure \ref{fig:endangeredCM} that the model has confidence in each of the 7 species. Some species such as the Rudds lark have much higher precision scores than species such as the Blue swallow. However it is important to note that the model does not confuse species. The decline of some species that the model has been trained on such as the Knysna Scrub-Warbler have been well documented in literature \cite{pryke2011persistence} these studies stand to benefit from more accurate sampling with this classification pipeline.

\section{Conclusion}
Through a series of experiments considering every aspect of the audio classification pipeline a few interesting insights were gained. It is evident that data augmentation does not provide good enough results to justify its usage. Transfer learning produced promising results, delivering better performance than a baseline model trained from scratch. All spectrogram models were able to retain their accuracy, and in some cases increase their performance metrics drastically after pruning and quantization, this also makes it possible to deploy these models on edge devices. This research exposes a few areas of biodiversity monitoring that warrant further research. One area is that of lifelong learning, it would be ideal to add new classes to trained models without having to retrain them, and the use of timbral features used along side a DNN should be investigated further is the models small size and fast inferencing times are ideal for biodiversity monitoring.

\printbibliography

@misc{xeno, title={canto}, url={https://www.xeno-canto.org/}, journal={xeno}}

@inproceedings{Mixup,
  title={Deep convolutional neural network with mixup for environmental sound classification},
  author={Zhang, Zhichao and Xu, Shugong and Cao, Shan and Zhang, Shunqing},
  booktitle={Chinese Conference on Pattern Recognition and Computer Vision (PRCV)},
  pages={356--367},
  year={2018},
  organization={Springer}
}

@article{Specaug,
  title={Specaugment: A simple data augmentation method for automatic speech recognition},
  author={Park, Daniel S and Chan, William and Zhang, Yu and Chiu, Chung-Cheng and Zoph, Barret and Cubuk, Ekin D and Le, Quoc V},
  journal={arXiv preprint arXiv:1904.08779},
  year={2019}
}

@inproceedings{Cutmix,
  title={Cutmix: Regularization strategy to train strong classifiers with localizable features},
  author={Yun, Sangdoo and Han, Dongyoon and Oh, Seong Joon and Chun, Sanghyuk and Choe, Junsuk and Yoo, Youngjoon},
  booktitle={Proceedings of the IEEE International Conference on Computer Vision},
  pages={6023--6032},
  year={2019}
}

@article{li2017hyperband,
  title={Hyperband: A novel bandit-based approach to hyperparameter optimization},
  author={Li, Lisha and Jamieson, Kevin and DeSalvo, Giulia and Rostamizadeh, Afshin and Talwalkar, Ameet},
  journal={The Journal of Machine Learning Research},
  volume={18},
  number={1},
  pages={6765--6816},
  year={2017},
  publisher={JMLR. org}
}

@inproceedings{hershey2017cnn,
  title={CNN architectures for large-scale audio classification},
  author={Hershey, Shawn and Chaudhuri, Sourish and Ellis, Daniel PW and Gemmeke, Jort F and Jansen, Aren and Moore, R Channing and Plakal, Manoj and Platt, Devin and Saurous, Rif A and Seybold, Bryan and others},
  booktitle={2017 ieee international conference on acoustics, speech and signal processing (icassp)},
  pages={131--135},
  year={2017},
  organization={IEEE}
}

@article{blalock2020state,
  title={What is the state of neural network pruning?},
  author={Blalock, Davis and Ortiz, Jose Javier Gonzalez and Frankle, Jonathan and Guttag, John},
  journal={arXiv preprint arXiv:2003.03033},
  year={2020}
}

@article{han2015deep,
  title={Deep compression: Compressing deep neural networks with pruning, trained quantization and huffman coding},
  author={Han, Song and Mao, Huizi and Dally, William J},
  journal={arXiv preprint arXiv:1510.00149},
  year={2015}
}

@inproceedings{denil2013predicting,
  title={Predicting parameters in deep learning},
  author={Denil, Misha and Shakibi, Babak and Dinh, Laurent and Ranzato, Marc'Aurelio and De Freitas, Nando},
  booktitle={Advances in neural information processing systems},
  pages={2148--2156},
  year={2013}
}

@article{salamon2017deep,
  title={Deep convolutional neural networks and data augmentation for environmental sound classification},
  author={Salamon, Justin and Bello, Juan Pablo},
  journal={IEEE Signal Processing Letters},
  volume={24},
  number={3},
  pages={279--283},
  year={2017},
  publisher={IEEE}
}

@article{pryke2011persistence,
  title={Persistence of the threatened Knysna warbler Bradypterus sylvaticus in an urban landscape: do gardens substitute for fire?},
  author={Pryke, James S and Samways, Michael J and Hockey, Philip AR},
  journal={African Journal of Ecology},
  volume={49},
  number={2},
  pages={199--208},
  year={2011},
  publisher={Wiley Online Library}
}

@article{simonyan2014very,
  title={Very deep convolutional networks for large-scale image recognition},
  author={Simonyan, Karen and Zisserman, Andrew},
  journal={arXiv preprint arXiv:1409.1556},
  year={2014}
}

@article{howard2017mobilenets,
  title={Mobilenets: Efficient convolutional neural networks for mobile vision applications},
  author={Howard, Andrew G and Zhu, Menglong and Chen, Bo and Kalenichenko, Dmitry and Wang, Weijun and Weyand, Tobias and Andreetto, Marco and Adam, Hartwig},
  journal={arXiv preprint arXiv:1704.04861},
  year={2017}
}

@article{szegedy2016inception,
  title={Inception-v4, inception-resnet and the impact of residual connections on learning},
  author={Szegedy, Christian and Ioffe, Sergey and Vanhoucke, Vincent and Alemi, Alex},
  journal={arXiv preprint arXiv:1602.07261},
  year={2016}
}
\vspace{12pt}

\end{document}